\documentstyle[12pt,aaspp4,psfig,amstex]{article}

\def\snr{G292.0+1.8}

\begin{document}

\title{\large \bf An X-Ray Point Source and Synchrotron Nebula Candidate in 
the Supernova Remnant G292.0+1.8}

\author{
C.M. Olbert\altaffilmark{1} and J.W. Keohane \\
North Carolina School of Science and Mathematics,
   1219 Broad St., Durham, NC 27705 \\
As submitted to the Astrophysical Journal Letters, May 26$^{th}$, 2001.}

\altaffiltext{1}{Address as of Fall 2001: Columbia Astrophysics Laboratory, 
	Columbia University, 550 West 120th Street, New York, NY 10027}

\begin{abstract}
	We present archival data from the {\it Chandra X-Ray Observatory} that
reveal a bright point source to the southeast of the center of the young supernova
remnant \snr\ that is coincident with the peak of highest radio surface brightness.
The mostly featureless spectrum of the point source at coordinates (J2000)
$\alpha$=11$^h$24$^m$39$^s.2$, $\delta$=-59\arcdeg16\arcmin19\arcsec.8
is well fit by a three-parameter absorbed model with one power-law and two 
blackbody components.  We also argue that the neutron star is surrounded by a
synchrotron wind nebula based off of the source's hard emission and high radio 
and X-ray luminosities, each corresponding to a canonical wind nebula spin-down 
power, $\dot{E}\sim{10}^{36}$\,erg\,s$^{-1}$.
\end{abstract}


\section{Introduction}\label{sec:intro}
	The young, oxygen-rich supernova remnant (SNR) \snr\ has been studied in the optical 
(Goss {\it et al.}~1979; van der Bergh 1979), infrared and radio (Braun {\it et al.}~1986, 
hereafter cited as BGCR86) and X-ray (Tuohy, Burton, \& Clark 1982; Hughes\& Singh 
1994) bands.  Despite this work, no pulses nor any evidence of a 
neutron star has been detected (e.g. Kaspi {\it et al.}~1996).  \snr\ is among the brightest 
of the galactic SNRs, displaying broad, filamentary 
structure, particularly through its center on arcminute and arcsecond scales.  
In the radio it is observed to display a ``pseudo-Crab'' 
morphology (van der Bergh 1979; BGCR86), while its X-ray morphology 
as shown by {\it Einstein} (Tuohy, Burton, \& Clark 1982) is confined to a 
center-filled ``bar'', which is resolved by {\it Chandra} as a filamentary band 
across the entire remnant.

	Recent {\it Chandra} data of \snr\ have revealed two intriguing point sources.  The
first is a faint, non-thermal object that is probably not associated with the remnant.  The
second source is a hard point source just below the central bar in \snr, offset to the 
southeast of the center of the remnant.  The location of this object corresponds to the 
contour of highest radio surface brightness (BGCR86), corresponding to a 
surface brightness of 1.08 Jy\,beam$^{-1}$ at 843 MHz.  Due to the low angular resolution 
of previous X-ray observations and the proximity of the point sources to a filament of the 
central bar, it is not surprising that these objects were only detected with the high angular 
resolution of {\it Chandra}.

	We present images and spectra of the point sources in the SNR \snr, arguing 
that the brighter source is a neutron star that is physically associated with the remnant.
Moreover, we argue that this source shows evidence of a synchrotron pulsar wind 
nebula (PWN) around it.  For reviews of SNRs, 
neutron stars, PWNe, and neutron star/SNR associations, see Jones {\it et al.}~(1998), 
Becker \& Pavlov (2001), Gotthelf (2001), and Helfand (1998), respectively.

\section{Observations and Analysis}\label{sec:obs}
\begin{figure}[tbh]
  \centerline{\hbox{\psfig{figure=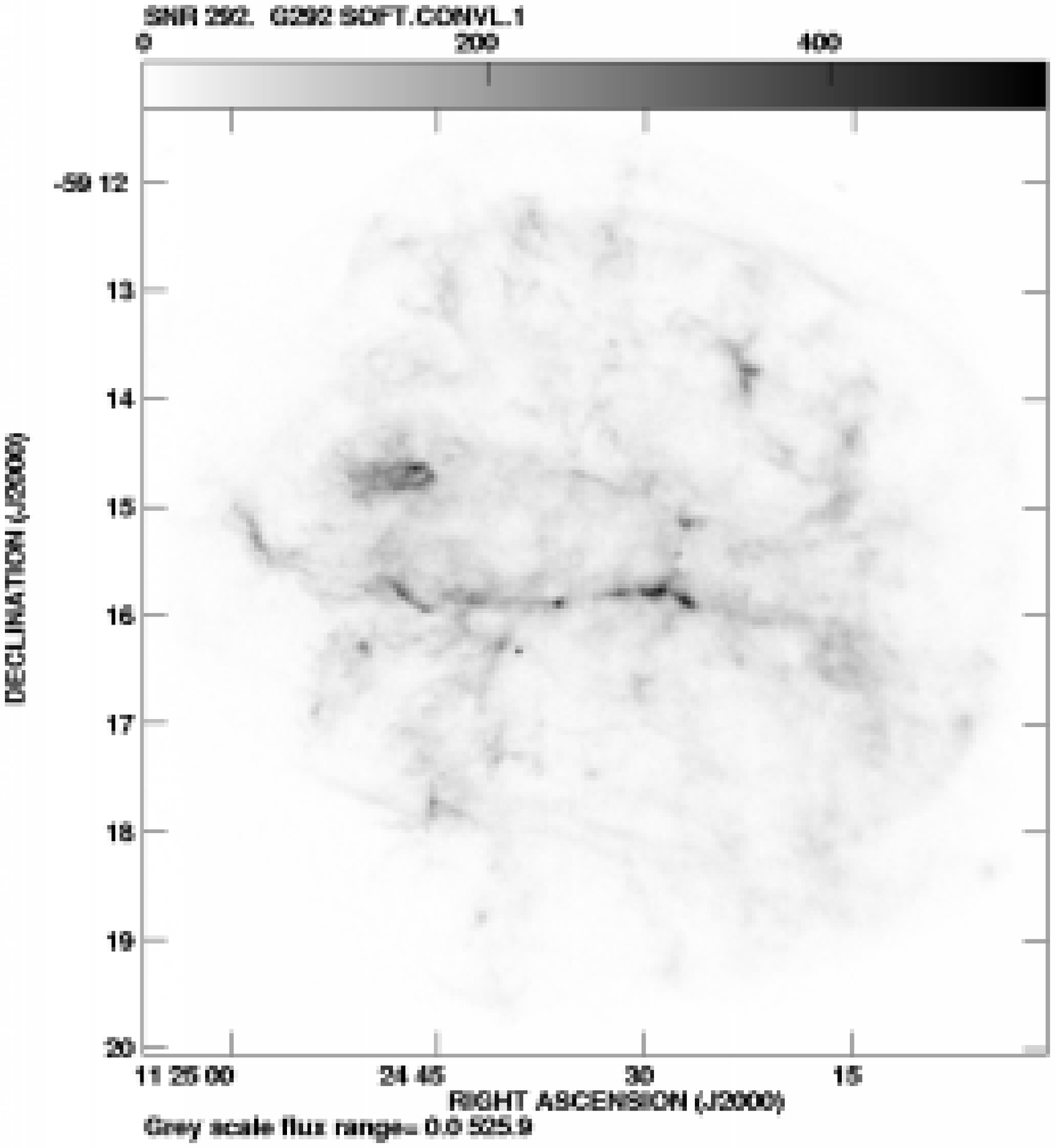,angle=0,width=7cm}}
{\psfig{figure=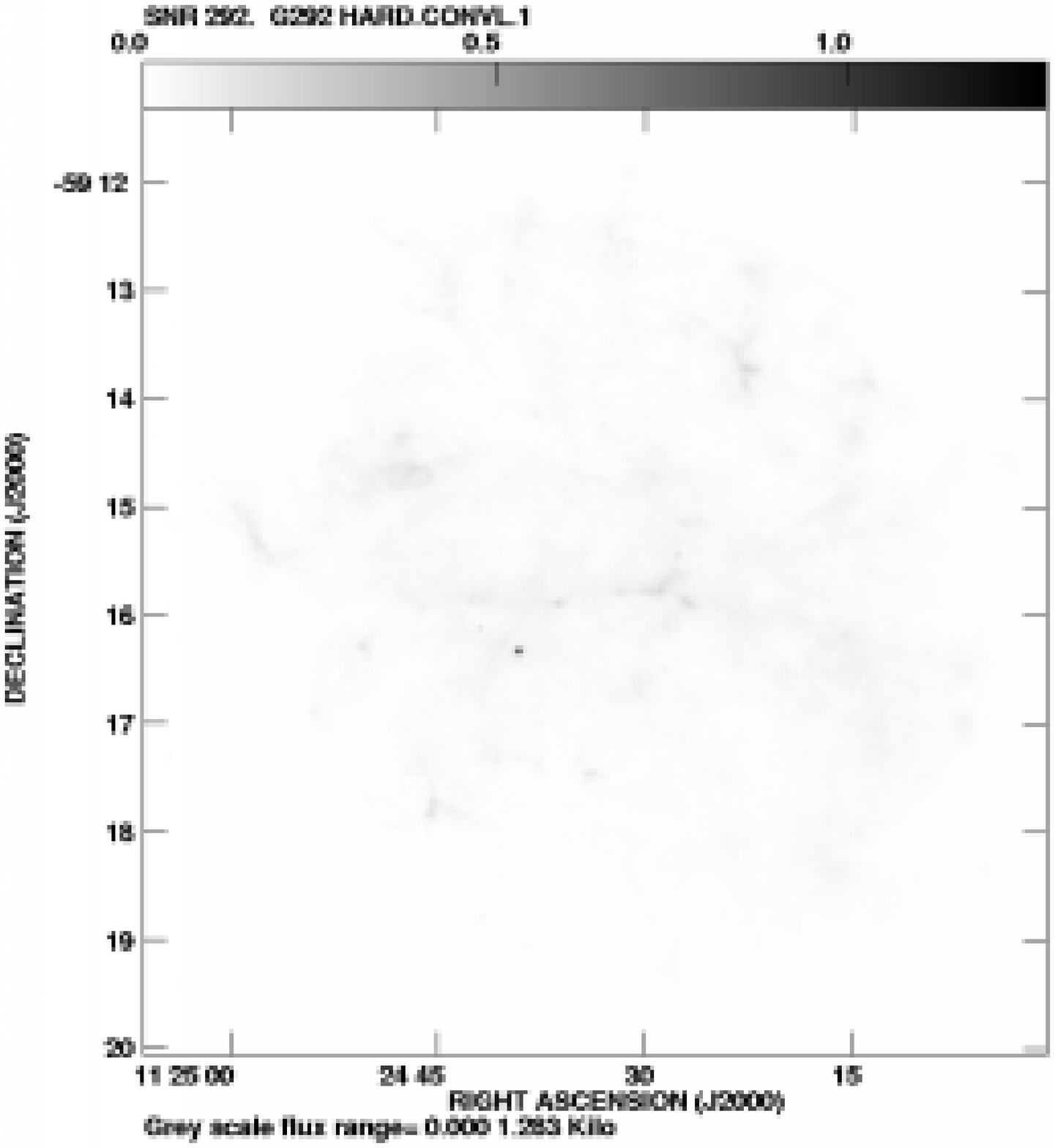,angle=0,width=7cm}}}
\caption[]{\small{Supernova remnant \snr\ shown in the soft (0.1-1.1\,keV) and hard 
(1.1-10.0\,kev) energy band, to the left and right respectively.  The filamentary, 
presumably thermal morphology of the remnant is particularly obvious in the soft 
image, while the point source almost exclusively stands out in the hard image.  Both 
images have been smoothed with a 2\arcsec\ beam.  The grayscale in the soft image 
ranges from 0.0\,counts/2\arcsec\ beam at the edge of the remnant to 
525.9\,counts/2\arcsec\ beam at the point source.  The grayscale in the hard image ranges 
from 0.0\,counts/2\arcsec\ beam at the edge of the remnant to 1283\,counts/2\arcsec\ beam 
at the point source.}}
\label{fig:remnant}
\end{figure}

\begin{figure}[tb]
  \centerline{\hbox{\psfig{figure=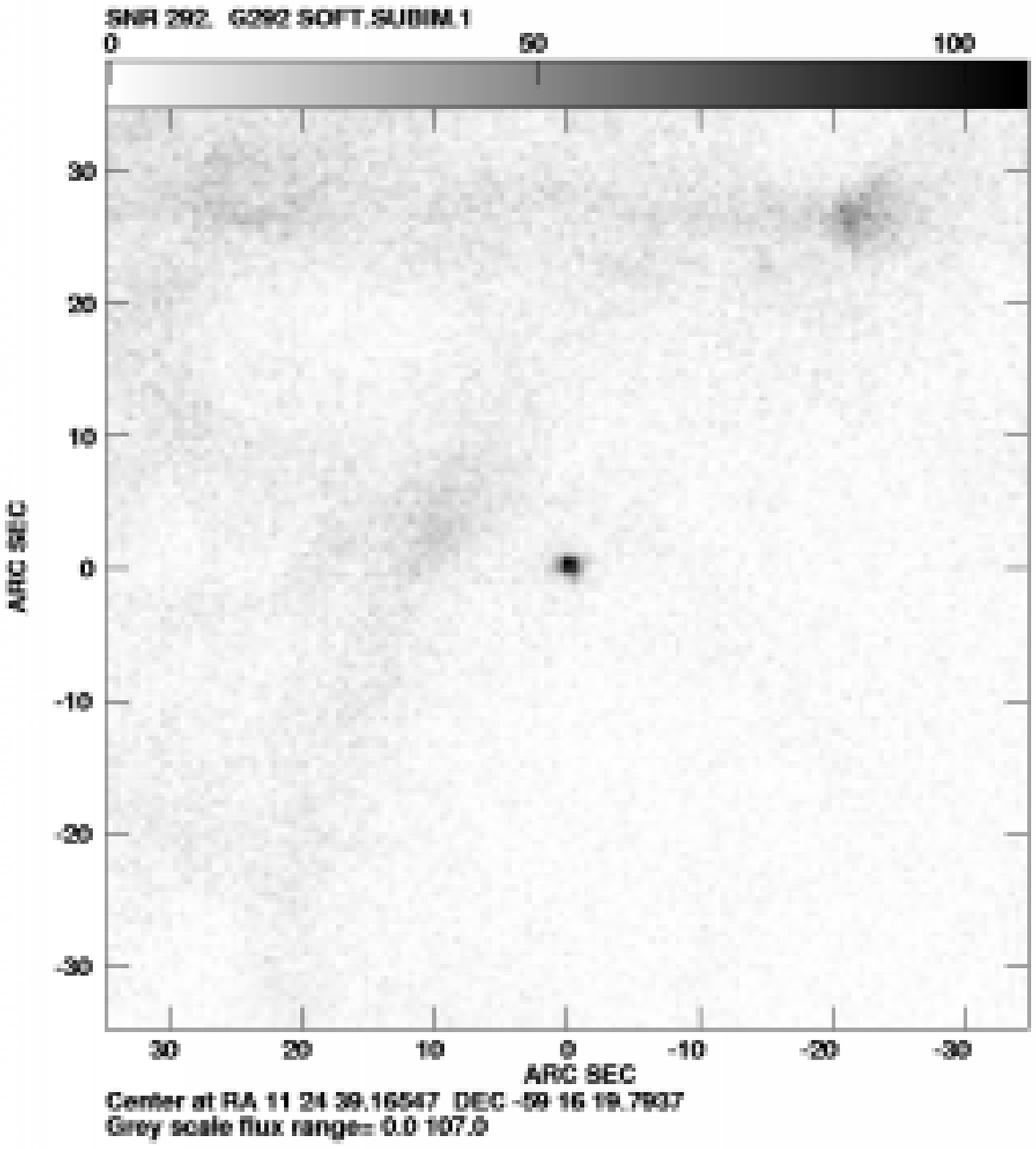,angle=0,width=7cm}}
{\psfig{figure=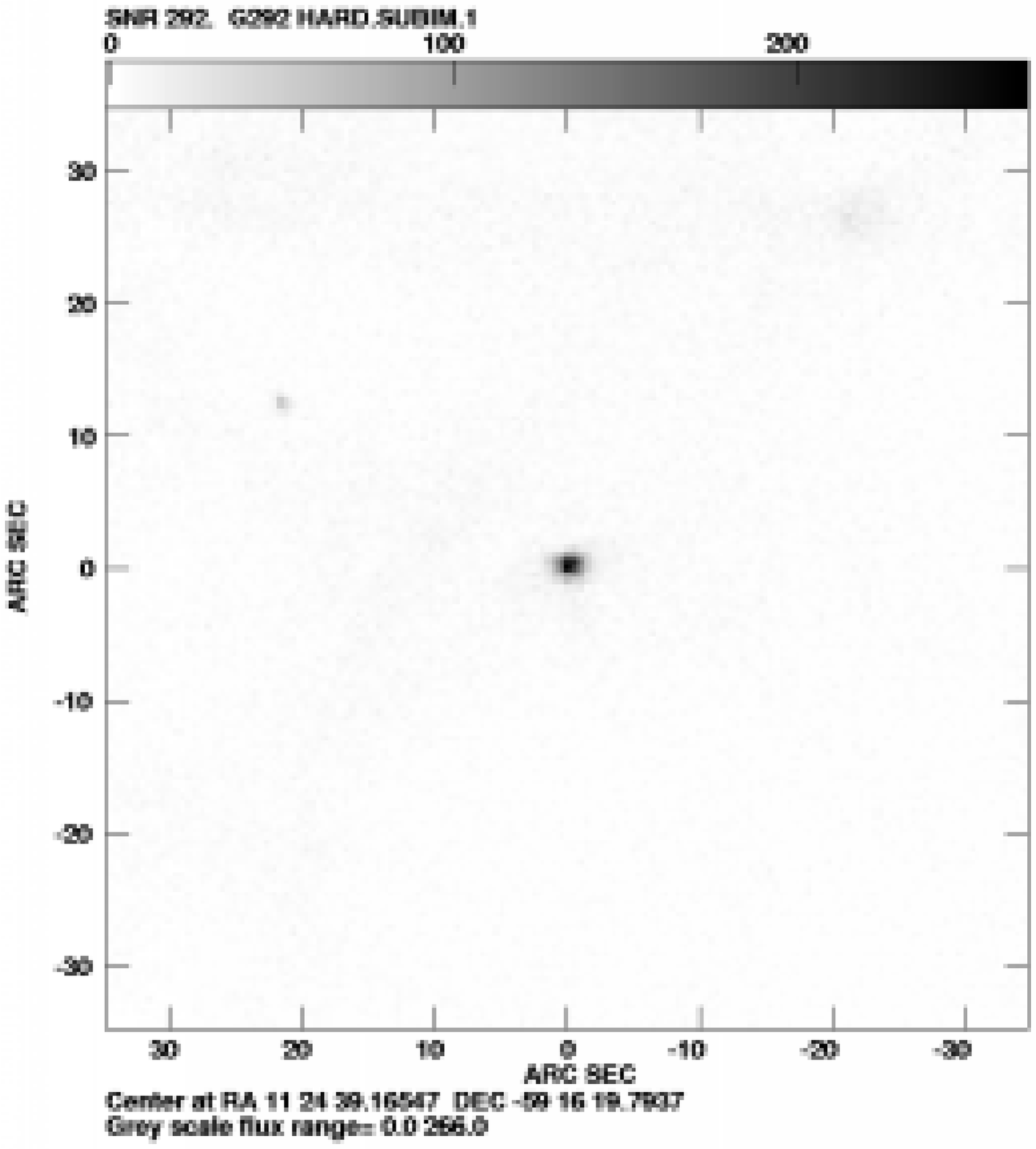,angle=0,width=7cm}}}
\caption[]{\small{A 70\arcsec x70\arcsec\ box centered at the brightest point source
in the soft (0.1-1.1\,keV) and hard (1.1-10.0\,keV) energy bands, to the left and
right respectively.  The center of the image is at coordinates (J2000)
$\alpha$=$11^h24^m39^s.2$, $\delta$=$-59\arcdeg 16\arcmin 19\arcsec.8$.  The
bright point source stands out plainly in both images, while the fainter point source
is only apparent in the hard energy band.  Neither image has been smoothed nor
binned.}}
\label{fig:zoom}
\end{figure}

\begin{figure}[tb]
  \centerline{\hbox{\psfig{figure=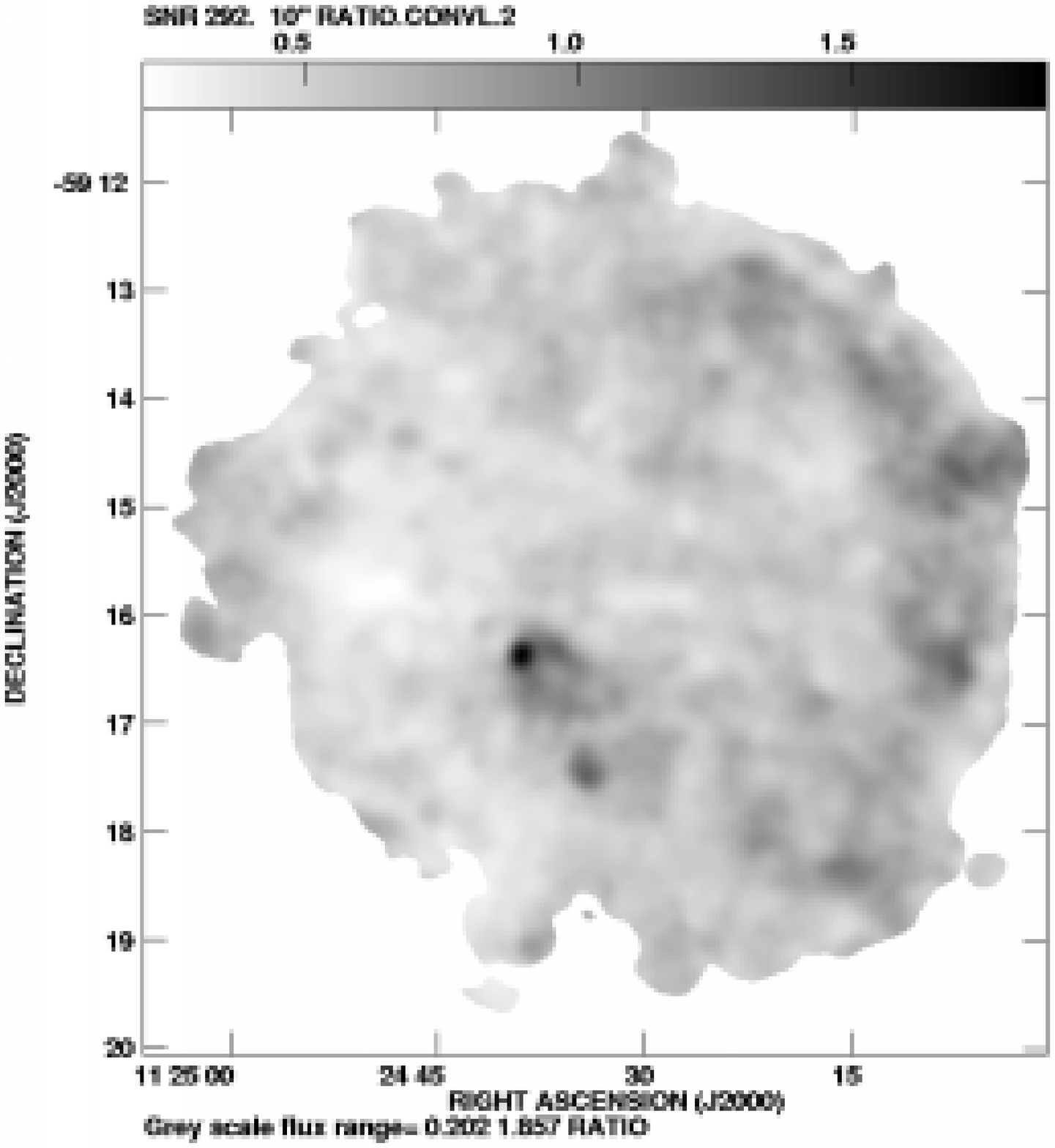,angle=0,width=7cm}}}
\caption[]{\small{Ratio map of the remnant: the hard (1.1-10.0\,keV) energy band divided by the 
soft (0.1-1.1\,keV) energy band images.  Each image was smoothed to 10\arcsec\
and cut at 10$\sigma$ before ratios were taken.  The fainter point source has been all but
washed out in the smoothing, while the brighter point source stands out dramatically against
the remnant background.  The dark extension to the southwest of the object is due to low
soft emission and not necessarily the presence of additional hard emission.  The grayscale 
range is from 0.202 to 1.857 ratio counts.}}
\label{fig:ratio}
\end{figure}

	{\it Chandra} performed a 40 ks GTO observation of \snr\ on March 11$^{th}$,
2000 with the Advanced CCD Imaging Spectrometer (ACIS).  The total emitting 
region of \snr\ fell onto the seventh chip in the ACIS array, CCD S-3.  The archival 
data were regained with CALDB v.2.3 and CIAO v.2.1.1 according to standard data 
processing procedures.

	A bright point source was detected at coordinates (J2000) {$\alpha$=11$^h$24$^m$39$^s.2$},
{$\delta$=-59\arcdeg16\arcmin19\arcsec.8}, which we have designated 
CXOU\,J112439-591619.  This source contains approximately 4500 
counts and is spread approximately normally in a ``circle'' of 4.75\,pixel=2.37\arcsec\ radius.  
A possible explanation for the partially-diffuse nature of the source is that there is a 
synchrotron pulsar wind nebula (PWN) around a neutron star, though there is no obvious 
morphological evidence of a bow-shock or synchrotron tail due to a high space velocity of 
the neutron star.  This possibility is discussed further in the following sections.

	Images were extracted of the entire remnant in the ``soft'' (0.1-1.1 keV) and 
``hard'' (1.1-10.0 keV) energy ranges (Fig \ref{fig:remnant}), with close-up images
surrounding the source included (Figure \ref{fig:zoom}).  This particular range for
the soft band was chosen based off of the knowledge that neutron stars primarily emit
soft, thermal X-rays in this range (Page 1995; Zavlin {\it et al.}~1995).  This object is 
particularly visible in the hard band image, which contains almost two-thirds of the total 
detected counts from the source.  Hardness ratio maps (hard image/soft image) were 
constructed, in which the point source stands out dramatically from the surrounding 
remnant (Figure \ref{fig:ratio}).

	Also standing out in the hardness ratio maps is a faint point source at coordinates (J2000)
$\alpha$=11$^h$24$^m$42$^s$, $\delta$=-59\arcdeg16\arcmin07\arcsec.2 which is
undetected in the soft band image.  Only $\sim$200 counts were detected towards 
this source, virtually all of which reside in the hard energy band.  No soft band emission
apart from typical remnant emission and background exists at these coordinates.  The 
object is quite visible in thehard-band image (Figure \ref{fig:zoom}) and the hardness
ratio map (Figure \ref{fig:ratio}).

\section{Spectra}\label{sec:spec}
	Spectra were extracted of the bright point source (a 2.37\arcsec\ radius 
circle centered at the point source emission) with an annulus of identical inner radius and 
a 22.1\arcsec\ radius subtracted as background.  The softest channel was
ignored, the PI channels were binned by a factor of 16, and the channels corresponding
to energies above 8.0\,keV were ignored to insure that there were at least 
10\,counts per bin.

\begin{deluxetable}{ccccccccc}
\tabcolsep0in\footnotesize
\tabcolsep0.1in
\tablewidth{\hsize}
\tablecaption{Spectral Fits of the Point Source \label{tab:spec}}
\tablehead
{
        \colhead{Model} &
        \colhead{N$_H^a$} &
        \colhead{kT$_1$} &
        \colhead{R$_{BB\,1}^b$} &
	\colhead{kT$_2$} &
	\colhead{R$_{BB\,2}$} &
	\colhead{$\Gamma$} &
	\colhead{Norm$_\Gamma$} &
	\colhead{$\chi_{\nu}^2$} \\
	\colhead{} & 
	\colhead{(10$^{22}$)} & 
	\colhead{(keV)} & 
	\colhead{(km)} & 
	\colhead{(keV)} & 
	\colhead{(km)} &
	\colhead{} & 
	\colhead{(10$^{-4}$)} &
	\colhead{}
}
\startdata
BB & 
	0.061$_{-0.01}^{+0.02}$ & 
	0.76$_{-0.02}^{+0.03}$ & 
	0.17 & 
	... & 
	... & 
	... & 
	... & 
	7.47 \\
&
	0.062$_{-0.01}^{+0.02}$ &
	0.72$_{-0.02}^{+0.03}$ &
	0.14 &
	... &
	... &
	... &
	... &
	5.89 \\
PL & 
	0.392$_{-0.028}^{+0.036}$ & 
	... & 
	... & 
	... & 
	... & 
	1.73$_{-0.1}^{+0.1}$ & 
	2.27 & 
	0.829 \\
&
	0.366$_{-0.024}^{+0.036}$ &
	... &
	... &
	... &
	... &
	1.78$_{-0.1}^{+0.1}$ &
	1.30 &
	0.763 \\
BB+BB & 
	0.198$_{-0.039}^{+0.046}$ & 
	0.40$_{-0.05}^{+0.05}$ & 
	0.392 & 
	1.39$_{-0.15}^{+0.24}$ & 
	.373 & 
	... & 
	... & 
	2.30 \\
&
	0.179$_{-0.033}^{+0.061}$ &
	0.39$_{-0.04}^{+0.09}$ &
	0.325 &
	1.5$_{-0.21}^{+0.25}$ &
	0.034 &
	... &
	... &
	1.03 \\
BB+PL & 
	0.425$_{-0.084}^{+0.187}$ & 
	0.16$_{-0.16}^{+0.02}$ & 
	1.28 & 
	... & 
	... & 
	1.72$_{-0.08}^{+0.03}$ & 
	2.11 & 
	0.802 \\
&
	0.425$_{-0.10}^{+0.25}$ &
	0.15$_{+0.02}^{+0.02}$ &
	1.73 &
	... &
	... &
	1.73$_{-0.09}^{+0.11}$ &
	1.26 &
	0.688 \\
BB+BB+PL & 
	0.425$_{-0.20}^{+0.28}$ & 
	0.16$_{-0.07}^{+0.02}$ & 
	1.75 & 
	0.51$_{-0.25}^{+0.15}$ & 
	0.110 & 
	1.61$_{-0.13}^{+0.12}$ & 
	1.74 & 
	0.855 \\
&
	0.426$_{-0.21}^{+0.31}$ &
	0.15$_{-0.01}^{0.02}$ &
	2.4 &
	0.47$_{-0.34}^{+0.07}$ &
	0.19 &
	1.18$_{-0.26}^{+0.21}$ &
	0.497 &
	0.655
\enddata
\tablecomments{\footnotesize{The first row for each model are the fit parameters
for the r=2.37\arcsec\ region centered at the point source, while the second row are
the fit parameters for the r=0.5\arcsec\ region (see \S\ref{sec:spec}).
\\
(a) - Frozen at 0.425, then unfrozen and fit when other
parameters pegged.  The value 0.425 was seen to be robust in all but the single 
and double blackbody fit.
\\ 
(b) - Assuming a distance to \snr\ of\,3.6 kpc (BGCR86).}}
\end{deluxetable}

\begin{figure}[tb]
  \centerline{\hbox{\psfig{figure=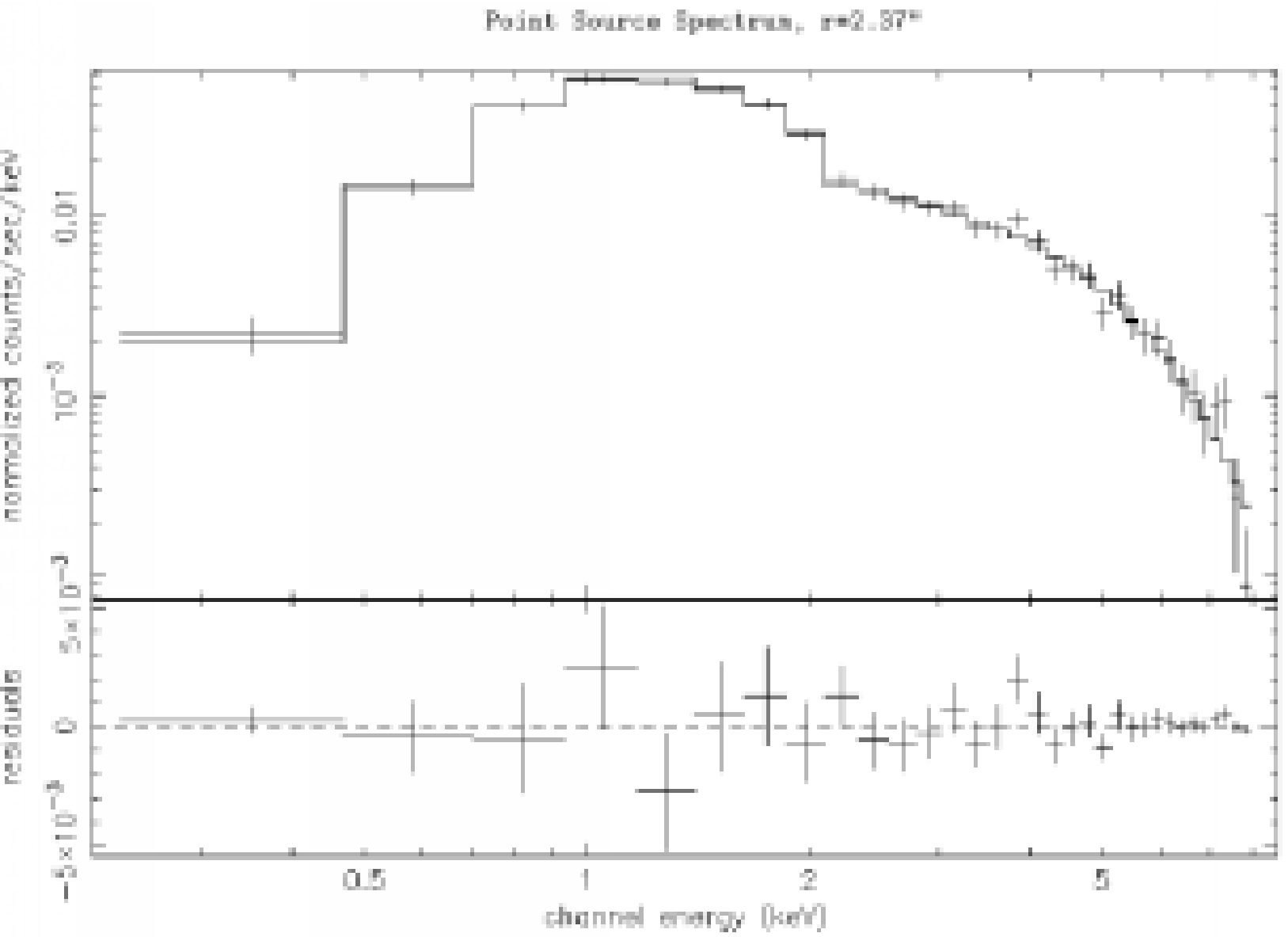,angle=0,width=8cm}}
{\psfig{figure=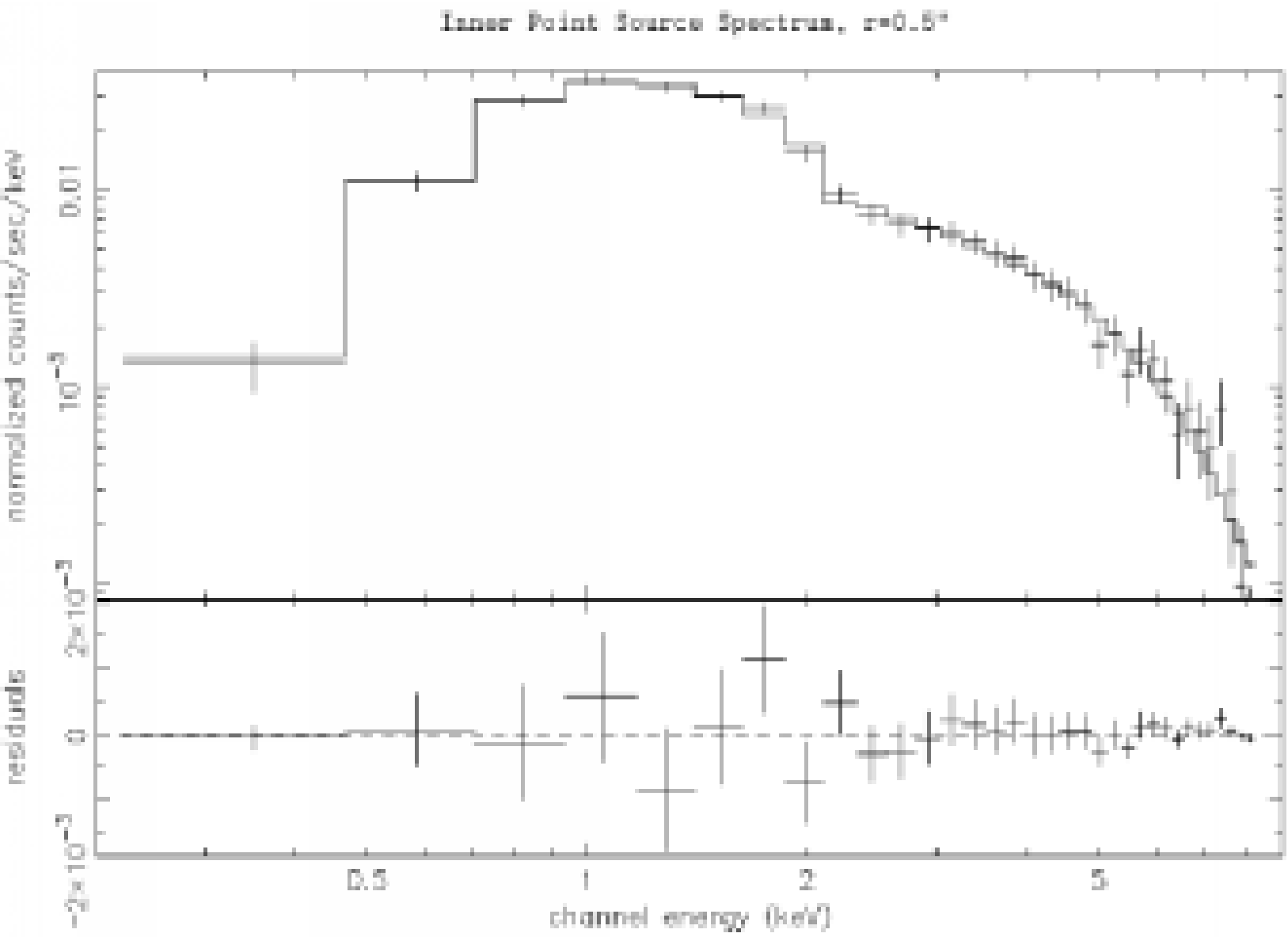,angle=0,width=8cm}}}
\caption[]{\small{Spectra of the entire point source (left, r=2.5\arcsec) and of the
reduced source (right, r=0.5\arcsec) with residuals below.  Note the difference in
line features above 3.0\,keV between the two spectra.  A model of 
wabs$\times$(bbodyrad$+$bbodyrad$+$power-law) is plotted on each spectrum.
The parameters for the each fit are outlined in Table \ref{tab:spec}.  A column density to
the source of N$_H$=0.425\,cm$^{-2}$ is robust in all of the statistically significant fits.}}
\label{fig:spec}
\end{figure}

	Attempts to separate neutron star emission from possible wind nebula emission were
ambiguous due to the compact nature of the source ($\sim$2.5\arcsec\ radius)
and due to sensitive background subtraction issues.  In our attempt, we used
a circle of $\sim$0.5\arcsec\ radius for the ``neutron star'' and an annulus of inner
$\sim$0.5\arcsec\ radius and outer 2.5\arcsec\ radius for the ``nebula''.  We also
subtracted a background annulus of 23\arcsec\ outer radius and identical
inner radius to the nebula (2.5\arcsec) as background.  The shapes of the inner circle and outer
annulus spectra were similar, with prominent spectral features above 3.0\,keV in the
annulus spectrum (Fig \ref{fig:spec}, \S\ref{sec:disc}).  A thorough discussion of the nebula 
spectrum seems to be unwarranted at this time, though we have included the inner 
circle spectrum in Figure \ref{fig:spec} and Table \ref{tab:spec} for comparison with our 
original fit of the r=2.37\arcsec\ spectrum.

	A variety of spectral models were fit to the point source spectrum in an attempt to 
find the most physically and statistically viable model of the emission.  The results of 
these spectral fits are summarized in Table \ref{tab:spec}.  The three-component spectral 
model blackbody radii correspond to luminosities of 
L$_1$=3.5$\times{10}^{32}$\,d$_{3.6}^2$\,erg\,s$^{-1}$ 
and  L$_2$=5.9$\times{10}^{31}$\,d$_{3.6}^2$\,erg\,s$^{-1}$, respectively, where d$_{3.6}$
is the distance to \snr\ in units of 3.6\,kpc.  Estimated X-ray fluxes (over the 0.2-4.0\,keV 
range) of F$_x$=5.0$\times{10}^{-13}$\,erg\,cm$^{-2}$\,s$^{-1}$ (for the entire source), 
and F$_x$=3.0$\times{10}^{-13}$\,erg\,cm$^{-2}$\,s$^{-1}$ (for the reduced radius source) 
give an X-ray luminosity on the order of
L$_x$=7.8$\times{10}^{32}$\,d$_{3.6}^2$\,erg\,s$^{-1}$.  We note that this
value of L$_x$ could be as high as 1.6$\times{10}^{33}$\,erg\,s$^{-1}$ or as low as
2.6$\times{10}^{32}$\,erg\,s$^{-1}$ given the error bars on d.  Plots of the 
compact source spectrum and the inner source spectrum mentioned above are shown in 
Figure \ref{fig:spec}.

	A spectrum of the fainter point source did not contain enough counts to provide
a statistically unique fit.  It is clear that the emission is almost entirely high-energy
($\ge$1.1\,keV), and shows evidence of emission and absorption lines above 2.0\,keV.  
It is statistically fit with a comparable column density to those
found for the bright source and a power-law of photon index $\Gamma$=1.2.  Thermal models
are excluded on both statistical and spectral grounds, as we find it difficult to believe that
such exclusively hard emission could be originating from a thermal source.  Blackbody, 
Raymond-Smith, Vpshock, Nei, and Sedov absorbed models either do not succeed in statistically
fitting the spectrum, or require disturbingly high temperatures on the order of kT$\sim$50-70\,keV.

\section{Discussion}\label{sec:disc}
	The bright point source in \snr\ appears to be one of a growing number of neutron 
star/supernova remnant associations (Helfand 1998), further study of which 
may ultimately provide us with clues as to how the composition of progenitor stars
and circumstellar material contribute to the evolution of neutron stars (e.g. Marsden
{\it et al.}~2001), and perhaps why these systems exist in such low abundance.
The fact that {\it Chandra} has begun to reveal a number of such point sources in
supernova remnants (e.g. Chakrabarty {\it et al.}~2001; Olbert {\it et al.}~2001) 
may indicate that the lack of neutron star/SNR associations is at least partly instrumental.

	Though the point source spectrum is statistically fit ($\chi_\nu\le$1) by power-law, 
blackbody plus power-law, and two blackbody plus power-law models, we argue that the latter 
is the most physically viable model.  We assert that the lower temperature kT=0.16\,keV 
(log\,T=6.27\,K) and larger blackbody radius represents the characteristic temperature of a
significant fraction of the neutron star surface or atmosphere (nominally, $\sim$5\%, though 
this is extremely dependent on temperature and temperature distribution) .  
This temperature is consistent with surface
temperatures predicted by standard cooling models (e.g. Page 1995; Page \& Sarmiento 1996), 
and is twice as high as the average effective temperature of other neutron stars 
(Slane \& Lloyd 1995).  However, the possibility of non-thermal continuum 
emission cannot be statistically ruled out at this time.  

	The second, higher temperature of kT=0.50\,keV and 
smaller blackbody radius could correspond to emission from a hot polar cap or a ``hot 
spot'' on either the surface or atmosphere due to temperature anisotropy resulting from 
internal convection or magnetic effects (e.g. Greenstein \& Hartke 1983; Pavlov {\it et al.}~1994).  
If this is true, these data indicate that the
luminosity contribution of small, hot regions of a neutron star is nearly as high as that of the
rest of the emitting area.  Given the luminous nature of the source, it may 
be fruitful to search for thermal pulses in the 0.1-1.1\,keV band (Pavlov {\it et al.}~1994; 
Page 1995; Zavlin {\it et al.}~1995).  We attribute the power-law component to non-thermal 
synchrotron emission from electrons accelerated by the high dipolar surface field of the 
neutron star.  Alternatively, a distribution of temperatures could be causing a power-law
spectrum, though if this were the case, our spectrum of the reduced radius source
would not require a more significant blackbody component and a less significant power-law
component, which our spectral fits indicate it does (see \ref{tab:spec}).

	A final intriguing feature of the compact object spectrum is the presence of spectral
features above 3.0\,keV.  Lines at 3.0-3.1\,keV and above 7.0\,keV seem to correspond to
highly ionized argon and iron, though they could also be due to background emission from
the SNR itself that was not successfully subtracted.  Interestingly, if one extracts a spectrum 
of 0.5\arcsec\ radius instead of 2.37\arcsec, high-energy features become noticeably different 
(see Figure \ref{fig:spec}).  This may indicate the presence of heavy elements being ionized 
in the outer atmosphere or nebula of the neutron star.

	We can estimate the object's transverse velocity assuming that the neutron star 
received a kinetic ``kick'' from the initial supernova explosion and has traveled ballistically
to its current location.  The {\it Chandra} image of \snr\ indicates a radius of $\le$250\arcsec, 
which corresponds to a radius of $\le$4.3\,d$_{3.6}^2$\,$t_{kyr}^{-1}$\,pc, where $t_{kyr}$ is 
the age of the remnant in terms of kiloyears (we adopt $t_{kyr}\sim$1: BGCR86).  An
estimate of the object's offset from the blast center (either geometric or from BGCR86) 
yields 60{\arcsec}$\pm$20\arcsec, and therefore a transverse velocity of
${\rm v}\sim{1000}$\,d$_{3.6}^2$\,$t_{1}^{-1}$\,$\theta_{60}$\,km\,s$^{-1}$, where 
$\theta_{60}$ is the object's angular distance from the blast center in terms of 60\arcsec.  
This is likely an upper limit, since spherical asymmetry probably places the blast center 
closer to the object.  Also, the presence of large-scale density gradients could alter this
result (e.g. Dohm-Palmer \& Jones 1996; Hnatyk \& Petruk 1999).  

	Though kick velocities of this magnitude are not unheard of, recent studies indicate that 
the mean kick velocity of neutron stars is about a fourth of this number (Hansen \& Phinney, 
1997).  This may indicate that the remnant is older than 1,000 years, perhaps having an 
age closer to the 2300\,yr derived by Agrawal \& Riegler (1979) or even older.  Another 
possibility is that the remnant is closer than we have assumed.  Without VLBI resolution,
a proper motion study seems unlikely to resolve this issue until anytime soon, since an age
of 1,000 years corresponds to a velocity of 0.06\arcsec\,yr$^{-1}$.

	Though the spatial features of the compact object and its surroundings provide
little evidence for a bow-shock and synchrotron tail, and the spectral analysis to the
same effect is ambiguous, it is nevertheless reasonable to presume that the extended 
region around the point source is in fact a synchrotron nebula.  Evidence of such a nebula
includes the hard nature of the source (e.g. Figures \ref{fig:remnant}, \ref{fig:ratio}), the 
spectral fits (Figure \ref{fig:spec}, \S\ref{sec:spec}), and the diffuse nature of the
compact source.  The previously derived value of L$_x$ gives a spin-down power of the 
neutron star $\dot{E}$=4.0$\times{10}^{35}$\,erg\,s$^{-1}$ (Seward \& Wang 1988).

	We can estimate a radio luminosity L$_R$ given that the source is a 1.08 Jy\,beam$^{-1}$
source at 843 MHz with a spectral index of -0.37, assuming that this object is the cause of the 
highest radio surface brightness on BGCR86's MOST map, to which it is coincidental.
Integrating this spectrum from 10 MHz to 100 GHz, we derive a radio luminosity of
L$_R={4.53}\times{10}^{32}$\,d$_{3.6}$\,erg\,s$^{-1}$.  Assuming that $\dot{E}\sim{10}^4\,$L$_R$
(Frail \& Scharringhausen 1997; Gaensler {\it et al.}~2000), our value of L$_R$ gives
$\dot{E}\sim{4.5}\times{10}^{36}$\,erg\,s$^{-1}$.  We adopt
$\dot{E}$=${10}^{36}$\,erg\,s$^{-1}$, well in the range of other young neutron 
stars (see Becker \& Pavlov 2001) and other neutron stars with surrounding PWNe 
(e.g. Predehl \& Kulkarni 1994; Frail {\it et al.}~1996; Olbert {\it et al.}~2001).  

	All $\gamma$-ray emitting pulsars have ratios of $\dot{E}_{33}$\,d$_{kpc}^{-2}\ge$0.5
(Nel {\it et al.} 1996), where $\dot{E}_{33}$ is $\dot{E}$ in terms of 10$^{33}$ erg\,s$^{-1}$, 
and so we suggest that it is plausible that this object is such a $\gamma$-ray source (for 
our derived $\dot{E}$, this ratio is approximately 100 times greater than the flux threshold of 0.5).

	As a final note, we interpret the faint, hard point source to the northwest of the asserted 
neutron star as a bright background object, perhaps an X-ray binary system.

\section{Conclusions}\label{sec:con}
	We have shown archival {\it Chandra} X-ray images and spectra of the
supernova remnant \snr.  These data have revealed a hard, bright X-ray point 
source at coordinates (J2000) $\alpha$=11$^h$24$^m$39$^s.2$, 
$\delta$=-59\arcdeg16\arcmin19\arcsec.8, which we have interpreted as an
isolated neutron star that is physically associated with the remnant.  We have
argued for a spectral model that includes one non-thermal and two thermal
components, and have obtained values similar to those of other neutron stars.
Likewise, the observed X-ray and radio luminosities and derived spin-down energy $\dot{E}$
are similarly consistent with other known neutron stars (e.g. Frail \&
Kulkarni 1991; Olbert {\it et al.}~2001).  We also assert that the diffuse nature of
the compact source (angular radius$\sim$2.5\arcsec) can be explained by the
presence of a synchrotron wind nebula around the source (Frail \& Kulkarni 1991; 
Frail {\it et al.}~1996).

	This object has presented a few problems that remain to be explained.  First
of all, the spectrally derived radius of the object is an order of magnitude smaller
than canonical values.  Also, our attempts to 
spatially or spectrally distinguish a synchrotron pulsar wind nebula have been 
ambiguous, though future observations of the object may resolve this issue.  Lastly, 
an estimate of the transverse velocity of the object is significantly higher than the
mean transverse velocity, though velocities of this magnitude are not unheard of.
Altogether, as with other new neutron star/SNR systems, the observation of
such systems pose more questions than they are able to answer.

\acknowledgements
We would like to thank C.R. Clearfield for his help with initial data analysis and
systems administration, and B.A. Pike \& N.E. Williams for their technical expertise.

\end{document}